\documentclass[aps,prl,showpacs]{revtex4}
\usepackage{epsfig}
\begin{document}
\title{A continuous time random walk model for financial distributions}
\author{Jaume Masoliver, Miquel Montero}
\address{Departament de F\'{\i}sica Fonamental, 
Universitat de Barcelona, Diagonal, 647, 08028-Barcelona, Spain}
\author{George H. Weiss}
\address{Center for Information Technology, National Institutes of Health, Bethesda, Md. 20892}

\date{\today}

\begin{abstract}
We apply the formalism of the continuous time random walk to the study of financial data. The entire distribution of prices can be obtained once two auxiliary densities are known. These are the probability densities for the pausing time between successive jumps and the corresponding probability density for the magnitude of a jump. We have applied the formalism to data on the US dollar/Deutsche Mark future exchange, finding good agreement between theory and the observed data. 
\end{abstract}
\pacs{89.65.Gh, 02.50.Ey, 05.40.Jc, 05.45.Tp}
\maketitle

\section{Introduction}
The continuous time random walk (CTRW), first introduced by Montroll and Weiss in 1965, \cite{montrollweiss}, has a large number of applications to the modelling of many physical phenomena, particularly in the field of transport in disordered media 
\cite{montroll2,weissllibre}. In contrast to the standard random walk in which steps are made periodically, the CTRW is based on the assumption that the times between steps is random. The CTRW has been applied in many different fieds. These range from transport in amorphous materials \cite{sher}, transport in turbid media \cite{wpm,gandj}, random networks \cite{berkowitz}, self-organized criticality \cite{boguna}, liquids \cite{berezhkovskii}, electron tunneling \cite{gudowska}, theoretical mechanics \cite{balescu} and earthquake modelling \cite{helmstetter}, just to name a few. 

In this paper apply CTRW formalism to a phenomenon more related to social sciences than to natural sciences: the distribution of speculative prices. The first analytical approach to this class of problems was proposed and analyzed by Bachelier who in 1900 modelled stock price movements as an ordinary random walk where prices can go up and down, at fixed times, due to a variety of many independent random causes. This approach necessarily leads to the conclusion that the probability distribution of speculative prices is Gaussian \cite{cootner}. In 1959 Osborne realized that, since stock prices are necessarily positive, it would be more convenient to consider returns instead of market values~\cite{osborne}. Thus, if $S(t)$ is an speculative price (or the value of an index) at time $t$ and
\begin{equation}
Z(t)=\ln[S(t)/S(0)]
\label{return}
\end{equation}
is the return up to time $t$. Then $Z(t)$ is a Gaussian variable and $S(t)$ is a log-normal process. Nevertheless, as Kendall first noticed in 1953 \cite{kendall}, the normal density fits financial data very poorly in the tails of the distribution. As an example: the probability of an event corresponding to five or more standard deviations is up to $10^4$ times larger than the one predicted by the Gaussian distribution. Therefore, empirical price distributions are highly leptokurtic. The existence of these ``fat tails" was precisely what led to Mandelbrot in 1963 to propose the L\'evy distribution for stock market prices \cite{mandelbrot}. There is, however, a drawback to this approach: no finite moments exist beyond the first and this is certainly a severe limitation of the model. Moreover, the L\'evy distribution has been tested against data in many situations, always with the same conclusion: the tails are far too long compared with actual data. In any case, as Mantegna and Stanley have recently shown \cite{mantegnastanley}, the L\'evy distribution fits very well to the center of empirical distributions ---surprisingly much better than the Gaussian density--- and it also shares the scaling behavior that appears in data. 

Recently a new market model was proposed to fill the gap between Gaussian and L\'evy distributions \cite{montero1}. The model, which was based on a continuous superposition of jump processes, explains the appearance of fat tails and self-scaling but still keeps all moments finite. It reproduces price distributions quite exactly, particularly those of tic-by-tic data. In this paper we want to address the problem from a different point of view. Thus we assume that the evolution of prices can be modeled by a CTRW. This allows us to calculate the distribution of speculative prices. The paper is organized as follows. In Sections II and III we set the general formalism and derive the exact distribution of prices and volatility. In Section IV we derive some asymptotic results mostly valid for long times. In Section V we apply the model to real data, namely the US dollar/Deutsche Mark future market. Conclusions are drawn in Section VI.

\section{General formalism. The distribution of prices}

\begin{figure}[htb]
\begin{center}
\includegraphics[width=12cm,keepaspectratio=true]{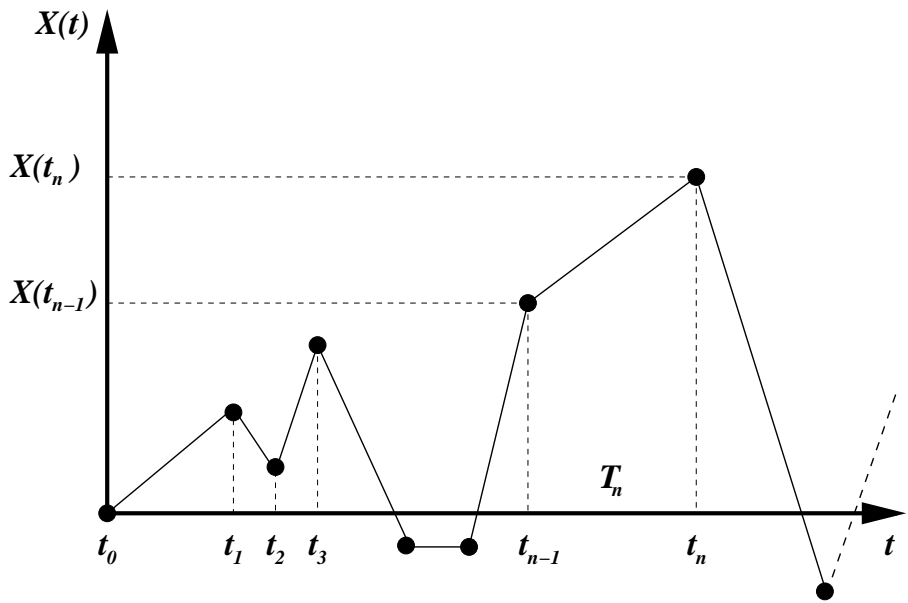}
\caption{Schematic representation of the return process. The dots mark the value $X(t_n)$ of the return after each sojourn. $T_n=t_n-t_{n-1}$ is the time increment of the n-th sojourn.}
\label{fig1}
\end{center}
\end{figure}

We define the zero-mean return $X(t)$ by
\begin{equation}
X(t)=Z(t)-\langle Z(t)\rangle,
\label{zeromeanreturn}
\end{equation}
where $Z(t)$ is given by Eq. (\ref{return}) and $\langle Z(t)\rangle$ is its average. We now suppose that $X(t)$ can be described in terms of a CTRW. In this picture $X(t)$ changes at random times $t_0,t_1, t_2,\cdots,t_n,\cdots$ and we assume that the intervals between successive steps (which we call ``sojourns") $T_n=t_n-t_{n-1}$ 
($n=1,2,3,\cdots$) are independent and identically distributed random variables with a probability density function given by $\psi(t)$, {\it i.e.,} 
$$
\psi(t)dt={\rm Prob}\{t<T_n\leq t+dt\}.
$$
At a given sojourn the zero-mean return $X(t)$ undergoes a random change giving rise to 
the random variable $\Delta X_n=X(t_n)-X(t_{n-1})$ which is described by a probability density function defined by
$$
h(x)dx={\rm Prob}\{x<\Delta X_n\leq x+dx\}.
$$
In this formulation of the problem we choose a function $\rho(x,t)$ to be the fundamental function, where $\rho(x,t)dxdt$ is the joint probability that an increment in return, $X(t)$, is added whose magnitude is between $x$ and $x+dx$ and that the time between successive turns is between $t$ and $t+dt$. The condition that there is no net drift will be assured by requiring that $\rho(x,t)$ is an even function of $x$. We can form two marginal densities out of $\rho(x,t)$: the pausing-time density $\psi(t)$ for the time between successive pulses
\begin{equation}
\psi(t)=\int_{-\infty}^{\infty}\rho(x,t)dx,
\label{psi}
\end{equation}
and the probability density function (pdf) for the changes in a single jump, $h(x)$, where 
\begin{equation}
h(x)=\int_{0}^{\infty}\rho(x,t)dt.
\label{h}
\end{equation}
The value of the return $X(t)$ at time $t$ will by given by the random value of the height at $t$ (see Fig. \ref{fig1}). We are interested in the probability density function of this variable, $p(x,t)$. 

We also assume that between successive steps the time evolution of $X(t)$ is linear (see Fig. \ref{fig1}). The assumption of linearity between jumps is arbitrary. We could have used step functions instead, in such a case the return would evolve  discontinuously and during any sojourn the value of the return (and hence the price) is that of the last jump. On the other hand, the linear choice for the return $X(t)$ implies an exponential growth in price $S(t)$ ({\it cf } Eq. (\ref{return})). This exponential behavior is an inherent feature of any financial market. 

Apparently the choice of linearity is against causality because one has to know future return values, $X(t_{n})$, in order to draw the segment joining $X(t_{n-1})$ and $X(t_{n})$. Fortunately for the soundness of our model this is not true. Indeed, notice that our process is completely equivalent to a random process evolving linearly with a slope changing at the random times $t_n$ $(n=1,2,3,\cdots)$. If the slope of the segment joining $X(t_{n-1})$ with $X(t_{n})$ is independent on $t_n, t_{n+1},\cdots$, then the random process is causal. Moreover, if the slope is independent on its previous values then the process is also Markovian. 

Let us calculate the form of $p(x,t)$ prior to the first jump. This function will be denoted by $p_0(x,t)$ and, due to the linear evolution of $X(t)$ between steps, it reads
\begin{equation}
p_0(x,t)=\int_t^\infty dT\int_{-\infty}^{\infty}\rho(y,T)
\delta\left(x-\frac{yt}{T}\right)dy.
\label{p0}
\end{equation}
where we have assumed that the initial jump occurred at $t=0$. In terms of $p_0$ and $\rho$ we have that the pdf $p(x,t)$ for the return at time $t$ is given by
\begin{equation}
p(x,t)=p_0(x,t)+\int_0^tdt'\int_{-\infty}^{\infty}\rho(x',t')p(x-x',t-t').
\label{inteq}
\end{equation}
This equation has been derived from the consideration that at time $t$, the process is either within the very first sojourn, this given by the first term on the rhs of 
Eq. (\ref{inteq}), or else the first sojourn ended at time $t'<t$, at which time the return had value $x'$, and from $(x',t')$ the process was renewed.

It is possible to solve Eq. (\ref{inteq}) by means of a joint Fourier-Laplace transform. To this end let us denote by 
$$
\hat{p}(\omega,s)=
\int_0^\infty dte^{-st}\int_{-\infty}^{\infty}d\omega e^{i\omega x}p(x,t), 
$$
the joint Fourier-Laplace transform of $p(x,t)$. Then the convolution theorems applied to Eq. (\ref{inteq}) yield
\begin{equation}
\hat{p}(\omega,s)=\frac{\hat{p}_0(\omega,s)}{1-\hat{\rho}(\omega,s)},
\label{hatp}
\end{equation}
where $\hat{p}_0(\omega,s)$ and $\hat{\rho}(\omega,s)$ are respectively the joint Fourier-Laplace transforms of functions $p_0(x,t)$ and $\rho(x,t)$. Recall that $p_0(x,t)$ only depends on $\rho(x,t)$, thus the return pdf $p(x,t)$ is exclusively determined by the form of $\rho(x,t)$. 

Unfortunately  the form of $\rho(x,t)$ is very difficult to determine from the available data. More easily accessible are the marginal densities of $\rho(x,t)$ in Eqs. (\ref{psi}) and (\ref{h}). It is therefore essential to assume a functional relation between $\rho(x,t)$ and its marginal densities $\psi(t)$ and $h(x)$. The simplest choice would be based on the assumption that return increments and their duration time are independent random variables. In this case
\begin{equation}
\rho(x,t)=h(x)\psi(t).
\label{independent}
\end{equation}
However, this assumption does not seem to be realistic since one certainly expects some degree of correlation between return increments and their duration, while 
Eq. (\ref{independent}) implies complete independence between increments and sojourn times. Following that intuition, for the rest of the paper we will mostly assume that the density $\rho(x,t)$ is such that its characteristic function $\tilde{\rho}(\omega,t)$ has the functional form:
\begin{equation}
\tilde{\rho}(\omega,t)=\psi\left[\frac{t}{\tilde{h}(\omega)}\right],
\label{fform}
\end{equation}
where $\psi(t)$ is the pausing time density, which we assume to be a decreasing function of time, \cite{lukacs}, and $\tilde{h}(\omega)$ is the characteristic function of $h(x)$. Notice that Eq. (\ref{fform}) has been chosen to satisfy Eqs. (\ref{psi}) and (\ref{h}). 

As we have mentioned, assumption given by Eq. (\ref{fform}) is at least intuitively plausible because it implies that one must wait for a long time in order for a large variation of return to occur. In other words, major increments of the return are very infrequent. We will prove this by showing that sojourn time $T=t_n-t_{n-1}$ and return quadratic increments $\Delta X^2=[X(t_n)-X(t_{n-1})]^2$ have a positive correlation, then increasing return variations imply increasing sojourn times and vice versa. In effect, we define the correlation function between $\Delta X^2$ and $T$ by
$$
r=\langle\Delta X^2 T\rangle-\langle\Delta X^2\rangle\langle\Delta T\rangle.
$$
We can easily evaluate the cross average $\langle\Delta X^2 T\rangle$ using the joint characteristic function $\tilde{\rho}(\omega,t)$.Thus
$$
\langle\Delta X^2 T\rangle=-\left.\frac{\partial^2}{\partial\omega^2}
\int_0^\infty t\tilde{\rho}(\omega,t)dt\right|_{\omega=0},
$$
which after using Eq. (\ref{fform}) yields
$$
\langle\Delta X^2 T\rangle=2\langle\Delta X^2\rangle\langle\Delta T\rangle.
$$
Hence
$$
r=\langle\Delta X^2\rangle\langle\Delta T\rangle>0,
$$
as we meant to prove.

Observe that assumption (\ref{fform}) allows us to write the joint Fourier-Laplace transform of $\rho(x,t)$ in the form
\begin{equation}
\hat{\rho}(\omega,s)=\tilde{h}(\omega)\hat{\psi}\left[s\tilde{h}(\omega)\right],
\label{hatrho}
\end{equation}
where $\hat{\psi}(s)$ is the Laplace transform of the pausing-time density $\psi(t)$. 
Likewise, from Eq. (\ref{p0}) we see that the transformed density $\hat{p}_0(\omega,s)$ can be written as
\begin{equation}
\hat{p}_0(\omega,s)=
-\frac{\partial}{\partial s}\int_0^1\frac{dz}{z}\hat{\rho}(\omega z,sz),
\label{hatp00}
\end{equation}
which, after using Eq. (\ref{hatrho}), reads
\begin{equation}
\hat{p}_0(\omega,s)=-\int_0^1\tilde{h}^2(\omega z)
\hat{\psi}'\left[sz\tilde{h}(\omega z)\right]dz,
\label{hatp0}
\end{equation}
where the prime denotes a derivative. In terms of the transformed densities $\hat{\psi}(s)$ and $\tilde{h}(\omega)$ the formal solution to the problem given by 
Eq. (\ref{hatp}) can be written in the following more explicit form
\begin{equation}
\hat{p}(\omega,s)=\frac{-\int_0^1\tilde{h}^2(\omega z)
\hat{\psi}'\left[sz\tilde{h}(\omega z)\right]dz}
{1-\tilde{h}(\omega)\hat{\psi}\left[s\tilde{h}(\omega)\right]}.
\label{formalsolution}
\end{equation}
Subject to the assumption in Eq. (\ref{fform}), Eq. (\ref{formalsolution}) furnishes a complete solution to the problem and it can be a convenient starting point for numerical methods when further analytical insight is unavailable. 

We finish this section with an example. Suppose that the random times $t_0,t_1,t_2,\cdots$ in which the return suffers random increments form a Poisson set of events, suppose also that these random increments are distributed according to a Laplace density. Then densities  $\psi(t)$ and $h(x)$ are respectively given by
\begin{equation}
\psi(t)=\lambda e^{-\lambda t},\qquad h(x)=\frac{\gamma}{2}e^{-\gamma|x|},
\label{poissoncase}
\end{equation}
where $\lambda^{-1}=\langle T\rangle$ is the mean sojourn time and $\gamma>0$ is such that $\langle\Delta X^2\rangle=2/\gamma^{2}$ is the jump variance. In this case $\tilde{h}(\omega)=1/(1+\omega^2/\gamma^2)$ and the characteristic function of the joint density $\rho(x,t)$ is given by
\begin{equation}
\tilde{\rho}(\omega,t)=\lambda\exp\left\{-\lambda(1+\omega^2/\gamma^2)t\right\},
\label{specialrho}
\end{equation}
which is the convolution of a Poissonian density $\lambda e^{-\lambda t}$ and a Gaussian density with zero mean and variance $2\lambda t/\gamma^2$. Finally, the inverse Laplace transform of the formal solution in Eq. (\ref{formalsolution}) now results in an explicit expression for the characteristic function of the problem given by the time convolution:
\begin{equation}
\tilde{p}(\omega,t)=\tilde{p}_0(\omega,t)+
\lambda\int_0^te^{-\lambda\omega^2t'/\gamma^2}\tilde{p}_0(\omega,t-t')dt',
\label{explicitsolution}
\end{equation}
where
\begin{equation}
\tilde{p}_0(\omega,t)=\lambda\int_t^\infty
\exp\left\{-\lambda\left(\tau+\frac{\omega^2t^2}{\gamma^2\tau}\right)\right\}d\tau.
\label{explicitp0}
\end{equation}
Note that numerical analyzing Eqs. (\ref{explicitsolution})-(\ref{explicitp0}) is straightforward.

\section{The volatility}

Besides the pdf $p(x,t)$, which provides all possible information about the problem, there is another quantity of considerable practical interest: the return variance. In our analysis this quantity, called ``volatility" in the terminology of finance, has the advantage that it does not require the knowledge of the entire jump distribution $h(x)$. It suffices to know the pdf  $\psi(t)$ and the first two moments of $h(x)$.

Let us denote by $\langle\widehat{X^n}(s)\rangle$ the Laplace transform of the $n$th moment of the return:
$$
\langle\widehat{X^n}(s)\rangle\equiv\int_0^\infty e^{-st}\langle X^n(t)\rangle dt.
$$
This can be written in terms of the joint Fourier-Laplace transform of $p(x,t)$ by
\begin{equation}
\langle\widehat{X^n}(s)\rangle=
i^{-n}\left.\frac{\partial^n\hat{p}(\omega,s)}{\partial\omega^n}\right|_{\omega=0}.
\label{moments1}
\end{equation}
Since we assume that there is no net drift in the evolution of $X(t)$, {\it i.e.}, $\rho(-x,t)=\rho(x,t)$, this means that all odd moments associated with $\rho(x,t)$ are zero. That is, 
\begin{equation}
\langle R^{2n-1}(t)\rangle=0,\qquad(n=1,2,3,\cdots),
\label{Rnodd}
\end{equation}
where
\begin{equation}
\langle R^m(t)\rangle\equiv\int_{-\infty}^{\infty}x^m\rho(x,t)dx
\label{Rn}
\end{equation}
is the $m$-th moment of the return increment. All of this means that random jumps during any sojourn are unbiased and, in particular, that their average is zero. Note that since the jump density $h(x)$ is a marginal density of $\rho(x,t)$, condition (\ref{Rnodd}) implies that
\begin{equation}
\tilde{h}^{(2n-1)}(0)=0,\qquad\mbox{ and}\qquad\mu_n\equiv(-1)^n\tilde{h}^{(2n)}(0),
\label{hprime}
\end{equation}
$(n=1,2,3,\cdots)$, where $\tilde{h}^{(m)}(0)$ is the $m$-th derivative of the characteristic function, $\tilde{h}(\omega)$, of the jump density. Notice that another direct consequence of the unbiased assumption given by Eq. (\ref{Rnodd}) is that all odd moments of the return process vanish:
\begin{equation}
\langle X^{2n-1}(t)\rangle=0,\qquad(n=1,2,3,\cdots).
\label{Xnodd}
\end{equation}

Starting from Eq. (\ref{moments1}) and using Eq. (\ref{hatp}) and 
Eqs. (\ref{Rnodd})-(\ref{Rn}) we obtain
\begin{equation}
\langle\widehat{X^2}(s)\rangle=\frac{\langle\widehat{X_0^2}(s)\rangle+
\langle\widehat{R^2}(s)\rangle}
{1-\hat{\psi}(s)},
\label{hatvol1}
\end{equation}
where $\langle\widehat{R^2}(s)\rangle$ and $\hat{\psi}(t)$ are the Laplace transforms of 
$\langle R^2(t)\rangle$ and $\psi(t)$ respectively, and 
\begin{equation}
\langle\widehat{X_0^2}(s)\rangle=
-\left.\frac{\partial^2\hat{p}_0(\omega,s)}{\partial\omega^2}\right|_{\omega=0}.
\label{hatvol01}
\end{equation}
The substitution of Eq. (\ref{hatp00}) into Eq. (\ref{hatvol01}) and some simple manipulations finally yield
$$
\langle\widehat{X_0^2}(s)\rangle=-\frac{1}{s}\langle\widehat{R^2}(s)\rangle+
\frac{2}{s}\int_0^1z\langle\widehat{R^2}(sz)\rangle dz,
$$
and Eq. (\ref{hatvol1}) implies
\begin{equation}
\langle\widehat{X^2}(s)\rangle=
\frac{2/s}{1-\hat{\psi}(s)}\int_0^1z\langle\widehat{R^2}(sz)\rangle dz.
\label{hatvol2}
\end{equation}

As we have mentioned, the independent model given by Eq. (\ref{independent}) cannot be   used for describing actual markets, but for the sake of completeness we will also give the general expression of the volatility associated with the model. This result will serve to illustrate an important point regarding the asymptotic behavior of the volatility which will be discussed nearly at the end of the next section. 

Using Eq. (\ref{independent}) we have
$$
\langle\widehat{R^2}(s)\rangle=\mu_2\hat{\psi}(s),
$$
and
\begin{equation}
\langle\widehat{X^2}(s)\rangle=
\frac{2\mu_2/s}{1-\hat{\psi}(s)}\int_0^1z\hat{\psi}(sz)dz,
\label{hatvolindependent}
\end{equation}
while for the dependent model exemplified by Eq. (\ref{fform}) and after using 
Eq. (\ref{hatrho}) we have
$$
\langle\widehat{R}^2(s)\rangle=\mu_2\frac{d }{ds}\left[s\hat{\psi}(s)\right].
$$
Then
\begin{equation}
\langle\widehat{X^2}(s)\rangle=\frac{2\mu_2/s}{1-\hat{\psi}(s)}
\left[\hat{\psi}(s)-\int_0^1 z\hat{\psi}(sz)dz\right].
\label{volatility1}
\end{equation}
Observe that both Eqs. (\ref{hatvolindependent}) and (\ref{volatility1}) depend only on the pausing-time density and the second moment of the jump $\mu_2$. In this case and for the example given in Eqs. (\ref{poissoncase})-(\ref{explicitp0}) we can explicitly write
\begin{eqnarray}
\langle X^2(t)\rangle=-\frac{2}{3\gamma^2}+\frac{2\lambda t}{\gamma^2}-
\frac{2}{3\gamma^2}(-1&+&2\lambda t+\lambda^2t^2)e^{-\lambda t}\nonumber\\
&+&\frac{2}{\gamma^2}\lambda^2t^2(1+\lambda t/3){\rm E}_1(\lambda t),
\label{explicitvolatility}
\end{eqnarray}
where ${\rm E}_1(\lambda t)$ is the exponential integral. Using both the short time behavior and the asymptotic behavior of ${\rm E}_1(\lambda t)$ \cite{mos} we can easily see that 
$$
\langle X^2(t)\rangle\sim-(2/\gamma^2)\lambda^2t^2\ln\lambda t,\qquad(t\ll\lambda^{-1}),
$$
and
$$
\langle X^2(t)\rangle\sim (2\lambda/\gamma^2)t,\qquad(t\gg\lambda^{-1}).
$$
In this particular case we thus observe an anomalous diffusion-like behavior at short times and a diffusion-like behavior at long times. We will see next that this is a general feature of the model rather than a peculiarity of the example given by Eq. (\ref{poissoncase}).

\section{Asymptotic results}

In this section we obtain some approximate results regarding the characteristic function and the volatility of the process. These results will mostly refer to the behavior of the probability distribution for large $t$ and large $x$. To this end we will assume that the Laplace transform of the pausing-time density, $\hat{\psi}(s)$, has the following series expansion
\begin{equation}
\hat{\psi}(s)=1+\sum_{n=1}^{\infty}a_ns^n+\sum_{n=0}^{\infty}b_ns^{\alpha+n-1},
\label{hatpsi}
\end{equation}
where $\alpha>1$ is a non integer number. Note that Eq. (\ref{hatpsi}) is a fairly general assumption, because when $b_n\equiv 0$ for all $n$ then all moments $\langle T^n\rangle$ of $\psi(t)$ exist. In such a case $a_n=(-1)^n\langle T^n\rangle/n!$. On the other hand if $\psi(t)$ only possesses $N\geq 1$ moments, then $N+1<\alpha<N+2$ and $a_n=(-1)^n\langle T^n\rangle/n!$  only for $n\leq N$. Using Eq. (\ref{hatpsi}) we see that $\hat{\rho}(\omega,s)$ given by Eq. (\ref{hatrho}) reads
\begin{equation}
\hat{\rho}(\omega,s)=\tilde{h}(\omega)
\left[1+\sum_{n=1}^{\infty}a_ns^n\tilde{h}^n(\omega)+
\sum_{n=0}^{\infty}b_ns^{\alpha+n-1}\tilde{h}^{\alpha+n-1}(\omega)\right],
\label{hatrho1}
\end{equation}
while $\hat{p}_0(\omega,s)$ is ({\it cf} Eq. (\ref{hatp0})) 
\begin{equation}
\hat{p}_0(\omega,s)=-\sum_{n=1}^{\infty}a_ns^{n-1}\phi_n(\omega)-
\sum_{n=0}^{\infty}(\alpha+n)b_ns^{\alpha+n-2}\phi_{\alpha+n-1}(\omega),
\label{hatp01}
\end{equation}
where
\begin{equation}
\phi_k(\omega)\equiv\int_0^1z^{k-1}\tilde{h}^k(\omega z)dz.
\label{phik}
\end{equation}

We want to obtain an asymptotic expansion of the pdf $p(x,t)$ valid for large $t$ and $|x|$. As is well known, the large $t$ behavior is equivalent to the small $s$ behavior of the Laplace transform. Similarly the large $|x|$ behavior correspond to the small $\omega$ behavior in the Fourier domain. Having this in mind from Eq. (\ref{hatrho1}) we have for $\alpha>3$:
$$
\frac{1}{1-\hat{\rho}(\omega,s)}=
\frac{1}{1-\tilde{h}(\omega)-a_1\tilde{h}^2(\omega)s+O(s^2)},
$$
or equivalently,
\begin{equation}
\frac{1}{1-\hat{\rho}(\omega,s)}=\frac{-1}{a_1\tilde{h}^2(\omega)s
\{1-[1-\tilde{h}(\omega)]/[a_1\tilde{h}^2(\omega)s]+O(s)\}}
\label{approx1}
\end{equation}
Recall that $\omega$ is also small and in this case $\tilde{h}(\omega)\simeq 1$. We now assume that, in spite of $s$ being small, the range of small values of $\omega$ that we will consider is such that 
$$
\frac{|1-\tilde{h}(\omega)|}{|a_1\tilde{h}^2(\omega)s|}\ll 1.
$$
We can thus expand the rhs of Eq. (\ref{approx1}) with the result
\begin{equation}
\frac{1}{1-\hat{\rho}(\omega,s)}\simeq\frac{-1}{a_1\tilde{h}^2(\omega)s}
\left[1+\frac{1-\hat{\rho}(\omega,s)}{a_1\tilde{h}^2(\omega)s}+O(s)\right].
\label{approx2}
\end{equation}
On the other hand from Eq. (\ref{hatp01}) we can write
\begin{equation}
\hat{p}_0(\omega,s)=-a_1\phi_1(\omega)+O(s),
\label{approx3}
\end{equation}
where $\phi_1(\omega)$ is given by Eq. (\ref{phik}). The substitution of Eqs. (\ref{approx2}) and (\ref{approx3}) into Eq. (\ref{hatp}) yields
\begin{equation}
\hat{p}(\omega,s)\simeq\frac{\phi_1(\omega)}{\tilde{h}^2(\omega)s}
\left[1+\frac{1-\hat{h}(\omega)}{a_1\tilde{h}^2(\omega)s}+O(s)\right],
\label{approx4}
\end{equation}
up to the leading order, Eq. (\ref{approx4}) yields
\begin{equation}
\hat{p}(\omega,s)\simeq\frac{\tilde{N}(\omega)}{a_1s^2}+O\left(\frac{1}{s}\right),
\label{hatp1}
\end{equation}
where $a_1=-\langle T\rangle$ is equal to the negative of the mean time between successive jumps and 
\begin{equation}
\tilde{N}(\omega)\equiv\frac{[1-\hat{h}(\omega)]\phi_1(\omega)}{\tilde{h}^4(\omega)}.
\label{G}
\end{equation}
Thus by virtue of Tauberian theorems the asymptotic expression for large $t$ of the characteristic function $\tilde{p}(\omega,t)$ will be given by the Laplace inversion of Eq. (\ref{hatp1}):
\begin{equation}
\tilde{p}(\omega,t)\sim-\tilde{N}(\omega)\frac{t}{\langle T\rangle},\qquad(t\rightarrow\infty).
\label{tildep1}
\end{equation}
In this asymptotic case the market volatility could be evaluated starting from 
Eq. (\ref{volatility1}) and then following the procedure just described which involves the use of Eq. (\ref{hatpsi}). Nevertheless, it turns out to be much simpler to directly evaluate $\langle X^2(t)\rangle$ using the asymptotic expression for $\tilde{p}(\omega,t)$ given by Eq. (\ref{tildep1}). We thus have 
\begin{equation}
\langle X^2(t)\rangle\sim\mu_2 t/\langle T\rangle \qquad(t\rightarrow\infty).
\label{volatility2}
\end{equation}
Therefore, in this regime the volatility grows linearly with time, which suggests a diffusion-like behavior of the model at long times. 

It is also interesting to calculate the behavior of volatility at short times. To this end we first note that as $t\rightarrow 0$ no shift in return has occurred with high probability. Consequently the probability density function $p(x,t)$ of the process is approximately given by $p_0(x,t)$:
$$
p(x,t)\simeq p_0(x,t),\qquad(t\rightarrow 0),
$$
and the approximate expression of the volatility is
$$
\langle X^2(t)\rangle\simeq\int_{-\infty}^{\infty}x^2p_0(x,t)dx,\qquad(t\rightarrow 0),
$$
which, after using Eq. (\ref{p0}) yields
\begin{equation}
\langle X^2(t)\rangle\simeq 
t^2\int_{t}^{\infty}\langle R^2(\tau)\rangle\frac{d\tau}{\tau^2},\qquad(t\rightarrow 0),
\label{x2approx}
\end{equation}
where $\langle R^2(t)\rangle$, the second moment of $\rho(x,t)$, is defined by Eq. (\ref{Rn}). In terms of the characteristic function $\tilde{\rho}(\omega,t)$ we have
$$
\langle R^2(t)\rangle=
-\left.\frac{\partial^2\tilde{\rho}(\omega,t)}{\partial\omega^2}\right|_{\omega=0},
$$
and from Eq. ({\ref{fform}) we get (see Eq. (\ref{hprime}))
$$
\langle R^2(t)\rangle=-\mu_2t\psi'(t). 
$$
Substituting this equation into Eq. (\ref{x2approx}) yields
$$
\langle X^2(t)\rangle\simeq-\mu_2 t^2
\int_{t/\langle T\rangle}^{\infty}\frac{\psi'(\xi\langle T\rangle)}{\xi}d\xi,
\qquad(t \rightarrow 0),
$$
where we have used dimensionless units in writing the integral. An integration by parts yields
$$
\int_{t/\langle T\rangle}^{\infty}\frac{\psi'(\xi\langle T\rangle)}{\xi}d\xi=
-\psi'(t)\ln(t/\langle T\rangle)-
\langle T\rangle\int_{t/\langle T\rangle}^{\infty}d\xi\psi''(\xi\langle T\rangle)\ln\xi,
$$
where we have assumed that $\psi'(t)$ decreases fast enough at infinity. It is easy to convince oneself that as $t\rightarrow 0$ the dominant term on the right hand side of this equation is the first one. Hence,
$$
\int_{t/\langle T\rangle}^{\infty}\frac{\psi'(\xi\langle T\rangle)}{\xi}d\xi\sim
-\psi'(t)\ln(t/\langle T\rangle).
$$
Finally,
\begin{equation}
\langle X^2(t)\rangle\sim\mu_2\psi'(0)t^2\ln(t/\langle T\rangle),
\qquad(t\ll\langle T\rangle).
\label{volatility3}
\end{equation}
Since $t^2|\ln(t/\langle T\rangle)|<t$ when $t\ll\langle T\rangle$, the volatility grows slower than normal diffusion at short times. Therefore, the model exhibits an anomalous diffusion-like behavior at short times. This peculiar behavior of the volatility, {\it i.e.,} anomalous diffusion at short times and ordinary diffusion at long times, is a characteristic feature of the model, and we will see next that there seems to be empirical evidence of such a behavior in real markets. 

It could be argued that the anomalous behavior of the volatility at short times is a spurious consequence of the form of $p_0(x,t)$ which, in turn, is the result of the arbitrary assumption that the time evolution of $\langle X^2(t)\rangle$ between successive steps is linear ({\it cf } Eq. (\ref{p0}) and Fig. \ref{fig1}). We will prove that this is not the case and that the anomalous behavior of volatility is, at least, a direct consequence of assumption (\ref{fform}) which relates return increments with their duration. In effect, suppose that the expression for $p_0(x,t)$ given by Eq. (\ref{p0}) is valid but that return increments and time intervals are independent random variables. Then from Eq. (\ref{independent}) we have
\begin{equation}
\tilde{\rho}(\omega,t)=\tilde{h}(\omega)\psi(t).
\label{independent2}
\end{equation}
Therefore $\langle R^2(t)\rangle=\mu_2\psi(t)$ and Eq. (\ref{x2approx}) reads
$$
\langle X^2(t)\rangle\simeq\mu_2 t\int_{1}^{\infty}\frac{\psi(t\tau)}{\tau^2}d\tau.
\qquad(t\rightarrow 0),
$$
For $t$ sufficiently small we may write
$$
\langle X^2(t)\rangle\simeq\mu_2\psi(0)t,\qquad(t\rightarrow 0),
$$
and the model presents a diffusion-like behavior at short time scales. Hence, it is the dependence between return increments and their duration the reason for the anomalous behavior of the model at short times. We finally note that this provides a test for the validity of assumption in Eq. (\ref{independent}) since if actual data do not support diffusion-like behavior at short times then the assumption of independence between increments and their duration is inaccurate. 

We finish this section obtaining the asymptotic long time behavior of the volatility for the independent model given by Eq. (\ref{independent}). In this case $\langle\widehat{X^2}(s)\rangle$ is given by Eq. (\ref{hatvolindependent}) which, for small $s$ and after using the expansion 
(\ref{hatpsi}), reads
$$
\langle\widehat{X^2}(s)\rangle\simeq
\frac{2\mu_2/\langle T\rangle}{s^2}
\left[\frac{1}{2}-\frac{1}{3}\langle T\rangle s+O(s^2)\right].
$$
Whence
$$
\langle X^2(t)\rangle\simeq\mu_2 t/\langle T\rangle,\qquad(t\rightarrow\infty)
$$
and, for the independent model, the volatility also has the same diffusion-like behavior at long time as that of the dependent model ({\it cf} Eq. (\ref{volatility2})).

\section{Specific results}

We will now apply the formalism presented in previous sections to analyze the distribution of returns corresponding to the future US dollar/Deutsche Mark exchange 
(tic-by-tic data from 1993 to 1997) \cite{futures,evertsz}. 

Before proceeding further we need to comment on a key point regarding the nature of markets. In a majority of works on the subject it is implicitly assumed that statistical properties of the economy are stationary all over the time. This is certainly inaccurate, for one does not expect current behavior of the market to be that of the market in, say, 1930. Here we presuppose a less restrictive assumption than that of complete stationarity: we suppose that markets are stationary over shorter periods of time, say one or two decades. Since we work on high frequency data and these data are only available since the early 1990's, our assumption of ``local (in time) stationarity" seems to be consistent with the data. 

As explained above, our first task is to infer from these data which forms for $\psi(t)$ and $h(x)$ are plausible. In Fig. \ref{fig2} we plot the experimental pausing-time density $\psi(t)$. We can see there that an excellent fit to the data is provided by the following pdf 
\begin{equation}
\psi(t)=\frac{\lambda(\alpha-1)}{(1+\lambda t)^{\alpha}},\qquad(3<\alpha<4),
\label{psifit}
\end{equation}
where $\alpha=3.47$ and $\lambda=2.73\times 10^{-2} s^{-1}$. Since $3<\alpha<4$ the first two moments of $\psi(t)$ are finite while the rest of moments do not exist. The Laplace transform of $\psi(t)$ is therefore of the class given by Eq. (\ref{hatpsi}). The mean sojourn time and the second moment are
\begin{equation}
\langle T\rangle=\frac{\lambda^{-1}}{\alpha-2},\qquad
\langle T^2\rangle=\frac{2\lambda^{-2}}{(\alpha-2)(\alpha-3)}.
\label{meantime}
\end{equation}
For the dollar/mark future market the experimental mean sojourn time evaluated from data is 
$\langle T\rangle_{exp}=23.65\ s$, in satisfactory agreement with the theoretical prediction of $\langle T\rangle=24.85\ s$ evaluated from Eq. ({\ref{meantime}) and based on ansatz (\ref{psifit}).

\begin{figure}[htb]
\begin{center}
\includegraphics[width=12cm,keepaspectratio=true]{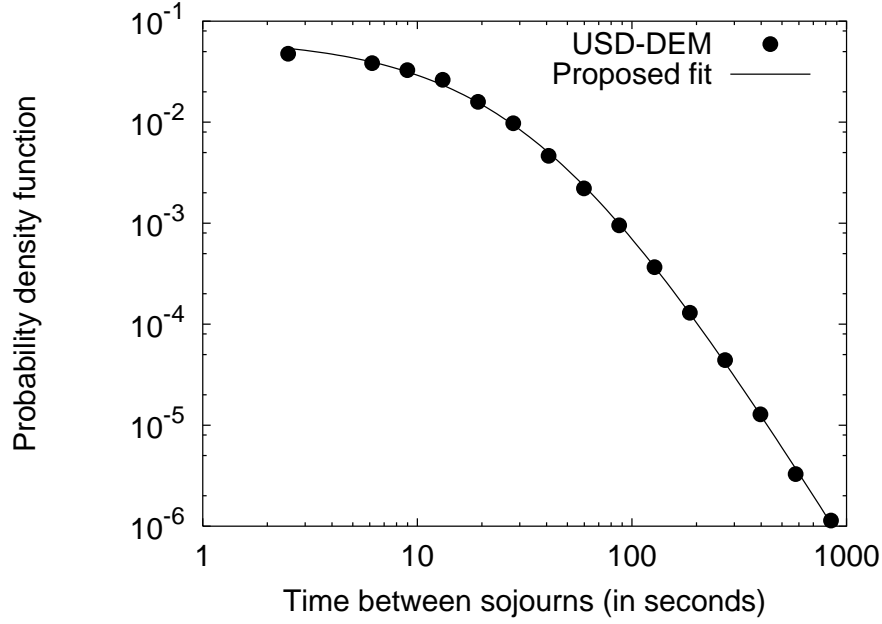}
\caption{Empirical distribution of the time between transactions, corresponding to the operative of the closest-to-maturity Deutsche Mark future (in the US market). The analyzed data range from January, 1993 to December, 1997. The pdf of the sojourn times, $\psi(t)$, clearly follows a power law. The solid curve represents the fit we propose in the main text.} 
\label{fig2}
\end{center}
\end{figure}

In Fig. \ref{fig3} we plot the experimental jump density $h(x)$. We see there that the experimental $h(x)$ can be considered a symmetric function of the return increments $x$. A  good fit is also given by a power law with a greater exponent than that of $\psi(t)$:
\begin{equation}
h(x)=\frac{(\beta-1)}{2\gamma(1+|x|/\gamma)^{\beta}},\qquad(5<\beta<6),
\label{hfit}
\end{equation}
where $\beta=5.52$ and $\gamma=2.64\times 10^{-4}$. Again, the pdf in (\ref{hfit}) has its four first moments finite and the rest are infinite. 

Power law densities like (\ref{psifit}) and (\ref{hfit}) have been recently suggested for describing several market models such as individual companies \cite{plerou} or market indices \cite{gropi2,mainardi}.

\begin{figure}[htb]
\begin{center}
\includegraphics[width=12cm,keepaspectratio=true]{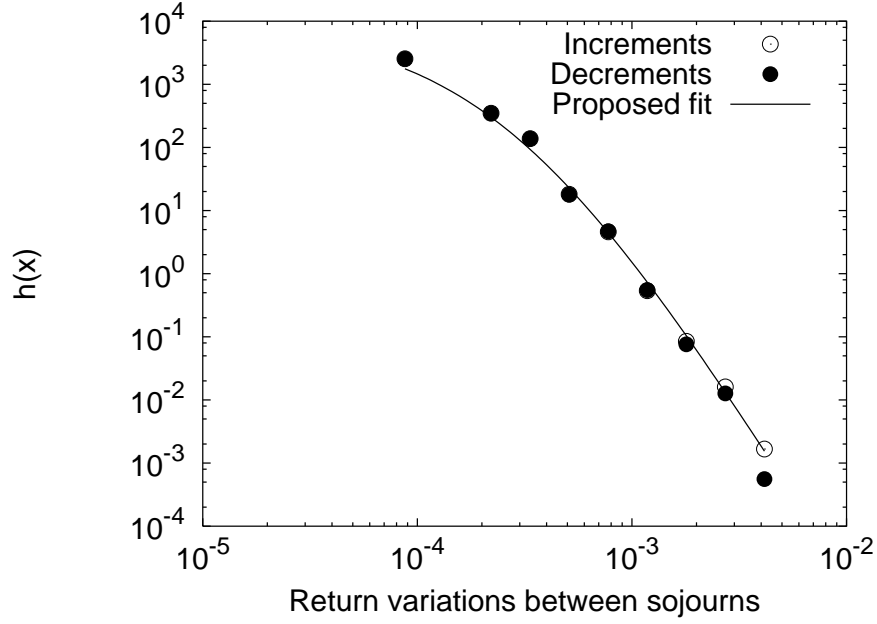}
\caption{Empirical distribution of the logarithmic changes between transactions, in the Deutsche Mark/US dollar future market. Positive variations (increments) and negative variations (decrements) exhibit approximately the same behavior, thereby supporting our assumption of the symmetry of $h(x)$. The plot also suggests the presence of a power law, and it includes a graph showing the shape of $h(x)$ in Eq.~(\ref{hfit}), using the parameters reported there.} 
\label{fig3}
\end{center}
\end{figure}

We now have all the ingredients to obtain a complete analysis of the dollar/mark exchange market. Unfortunately the densities given by Eqs. (\ref{psifit}) and (\ref{hfit}) make very problematic the exact evaluation of $p(x,t)$ by means of the Fourier-Laplace inversion of 
Eq. (\ref{formalsolution}). We will use instead the approximate results obtained in Sect. IV, and in particular the asymptotic expression Eq. (\ref{tildep1}) for the characteristic function $\tilde{p}(\omega,t)$ valid for large $t$. To this end we must have an expression for the jump characteristic function,
\begin{equation}
\tilde{h}(\omega)=2\int_0^\infty h(x)\cos\omega x dx.
\label{tildeh}
\end{equation}
Substituting Eq. (\ref{hfit}) into this equation enables us to obtain the exact  $\tilde{h}(\omega)$ in terms of a combination of incomplete gamma functions of imaginary arguments. However that expression is clumsy for practical purposes and, since we are mainly interested in the behavior of the tails of $p(x,t)$, we will use a simpler expression for $\tilde{h}(\omega)$ valid when $\omega$ is small. 

We therefore define
$$
\tilde{g}(\omega)=1-\tilde{h}(\omega)
$$
then, using Eqs. (\ref{hfit})-(\ref{tildeh}) and taking into account both normalization and  symmetry of $h(x)$, we write
$$
\tilde{g}(\omega)=2\int_0^\infty[1-h(x)]\cos\omega x dx.
$$
Substituting Eq. (\ref{hfit}) into this equation and recalling that $5<\beta<6$, after successive integrations by parts we find
\begin{eqnarray*}
\tilde{g}(\omega)=\frac{\Gamma(\beta-3)}{\Gamma(\beta-1)}(\gamma\omega)^{2}&-&
\frac{\Gamma(\beta-5)}{\Gamma(\beta-1)}(\gamma\omega)^{4}\\
&+&\frac{\Gamma(\beta-5)}{\Gamma(\beta-1)}(\gamma\omega)^{\beta-1}
\int_0^\infty\frac{\sin x}{x^{\beta-5}(1+\omega\gamma/x)^{\beta-5}} dx.
\end{eqnarray*}
As $\omega\rightarrow 0$ we make the approximation
$$
\int_0^\infty\frac{\sin x}{x^{\beta-5}(1+\omega\gamma/x)^{\beta-5}} dx\simeq
\int_0^\infty\frac{\sin x}{x^{\beta-5}}dx=\frac{\pi}{2\Gamma(\beta-5)\sin\pi(\beta-5)/2}.
$$
Hence
\begin{eqnarray}
\tilde{h}(\omega)\simeq 1-\frac{\Gamma(\beta-3)}{\Gamma(\beta-1)}(\gamma\omega)^{2}
&+&\frac{\Gamma(\beta-5)}{\Gamma(\beta-1)}(\gamma\omega)^{4}\nonumber\\
&-&\frac{\pi}{2\Gamma(\beta-1)\sin\pi(\beta-5)/2}(\gamma\omega)^{\beta-1}.
\label{tildeh1}
\end{eqnarray}
The Fourier inversion of this approximation will give us the behavior of $h(x)$ as $x\rightarrow\pm\infty$. Then, neglecting delta function terms (which obviously do not contribute for large values of $|x|$), we have
$$
h(x)\sim-\frac{1}{2\Gamma(\beta-1)\sin\pi(\beta-5)/2}\gamma^{\beta-1}
\int_0^{\infty}\omega^{\beta-1}\cos\omega x d\omega,
$$
$(x\rightarrow\pm\infty)$.The integral appearing on the right hand side of this equation is convergent in the sense of generalized functions and reads \cite{lighthill}
\begin{equation}
\int_0^{\infty}\omega^{\beta-1}\cos\omega x d\omega=
-\frac{\Gamma(\beta)\sin\pi(\beta-5)/2}{|x|^{\beta}}.
\label{lighthill}
\end{equation}
Therefore,
\begin{equation}
h(x)\sim\frac{\beta-1}{2}\frac{\gamma^{\beta-1}}{|x|^{\beta}},\qquad(|x|\rightarrow\infty),
\label{tailh}
\end{equation}
and the tails of the jump distribution follow a power law with exponent $\beta$. This is consistent with Eq. (\ref{hfit}) which in turn proves the soundness of the approximation in  Eq. (\ref{tildeh1}).  

\begin{figure}[htb]
\begin{center}
\includegraphics[width=12cm,keepaspectratio=true]{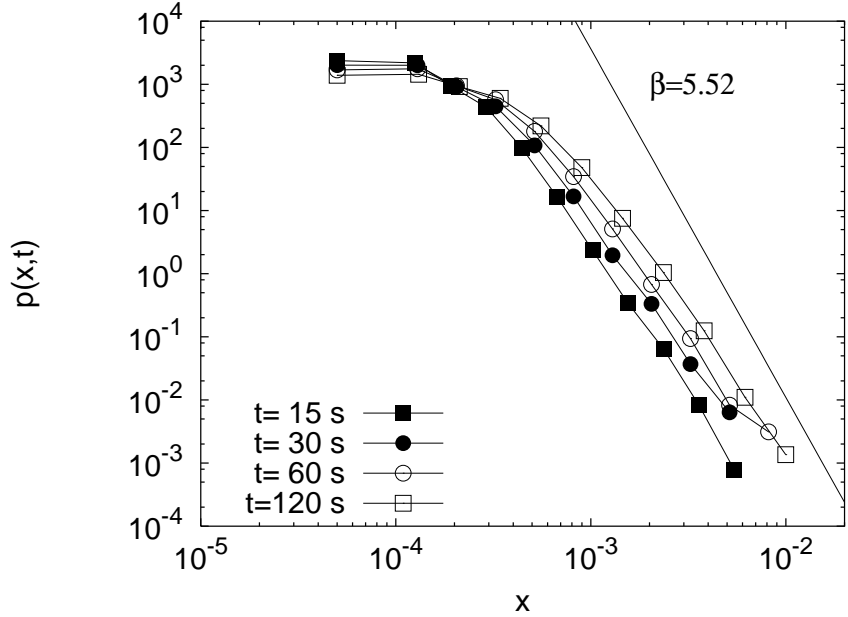}
\caption{Empirical probability density function $p(x,t)$ for a set of time lags $t$, ranging from 15 seconds to 2 minutes. The model leads to a power-law decay, governed by an exponent $\beta$, in all cases. This exponent is precisely the one appearing in the power-law for the jump density $h(x)$. Since we have chosen exponentially growing values for $t$, different times correspond to equally spaced tails and this indicates a linear growth with time. We only show tails for positive increments, tails for negative increments behave in the same way.}
\label{fig4}
\end{center}
\end{figure}

Let us now prove that the tails of the entire distribution $p(x,t)$ also obey a power law with the {\it same exponent} $\beta$ for any time $t$ sufficiently large. Indeed, the substitution of Eq. (\ref{tildeh1}) into Eq. (\ref{G}) yields
\begin{equation}
\tilde{N}(\omega)\simeq M(\gamma\omega)^{\beta-1},
\label{G1}
\end{equation}
where 
\begin{equation}
M=\frac{\pi}{2\Gamma(\beta-1)\sin\pi(\beta-5)/2}.
\label{M}
\end{equation}
In writing Eq. (\ref{G1}) we have taken into account the fact mentioned above that integer powers of $\omega$ do not affect the behavior at the tails. In this situation the asymptotic expression of $\tilde{p}(\omega,t)$ given by Eq. (\ref{tildep1}) can be written as 
\begin{equation}
\tilde{p}(\omega,t)\sim-Mt(\gamma\omega)^{\beta-1}/\langle T\rangle,
\label{tildep2}
\end{equation}
and the Fourier inversion of Eq. (\ref{tildep2}) finally reads \cite{lighthill}
\begin{equation}
p(x,t)\sim\frac{(\beta-1)t}{2\langle T\rangle}\frac{\gamma^{\beta-1}}{|x|^{\beta}},
\qquad(|x|\rightarrow\infty).
\label{tailp}
\end{equation}
Hence tails of $p(x,t)$ decay following the same power law as that of the return increment distribution $h(x)$. This prediction of the theoretical model is confirmed by actual data. In Fig. \ref{fig4} we show the empirical $p(x,t)$ for the dollar/mark future exchange and for different values of time $t$, ranging from 15 seconds to 2 minutes. The empirical distribution clearly shows, for all these times, a power-law decay with exponent $\beta\approx 5.5$, which coincides with the decaying exponent of $h(x)$ thus confirming the predictions of the CTRW model. Moreover, Eq. (\ref{tailp}) predicts the linear growth of tails with time. This linear growth is indeed observed in Fig. \ref{fig4} where different times correspond to properly spaced curves.

\begin{figure}[htb]
\begin{center}
\includegraphics[width=12cm,keepaspectratio=true]{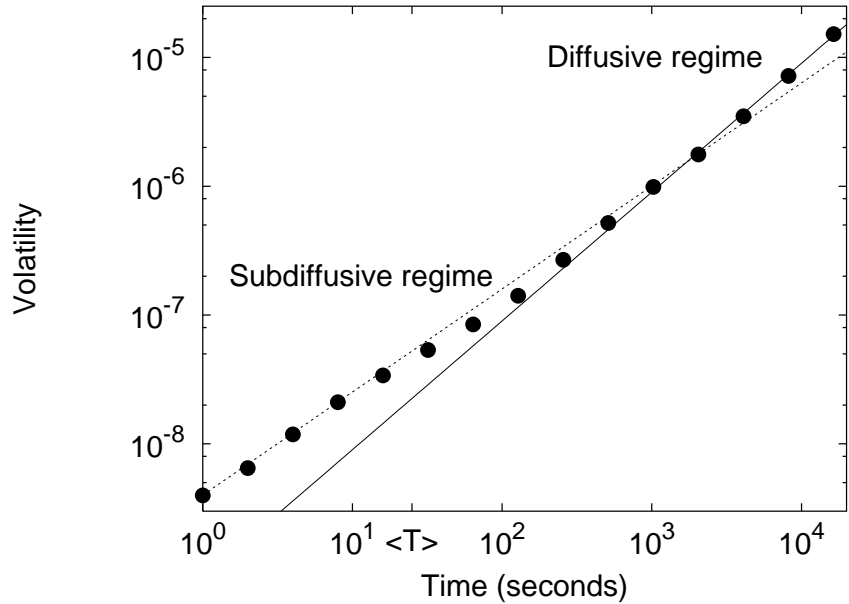}
\caption{The experimental volatility (dots) shows two different regimes. At short times there is a subdiffusion-like behavior (dotted line) while at long time the volatility grows linearly with time (solid line). Transition between regimes occurs approximately at time $t\simeq\langle T\rangle$. These facts are in agreement with theoretical predictions.} \label{fig5}
\end{center}
\end{figure}

Let us finally and briefly comment on the volatility. In Fig. \ref{fig5} we plot the volatility $\langle X^2(t)\rangle$ for the dollar/mark future market. We can see there that the experimental volatility clearly shows two different regimes: at short times we observe a subdiffusion-like behavior while at long times the volatility clearly appears to have a diffusion-like behavior. Both regimes are consistent with the CTRW dependent model. We also note from Fig. \ref{fig5} that the transition between these two regimes occurs at $t\simeq 25\ s$ which is approximately equal to the mean sojourn time $\langle T\rangle$.  Again this transition from subdiffusion to diffusion around time $t\simeq\langle T\rangle$ has been predicted by the dependent model.

\section{Conclusions}

In this paper we have applied formalism based on the CTRW to the random movement of market prices. The formalism depends on the conjecture from data of two densities: the pausing-time density $\psi(t)$ and the jump density $h(x)$. The assumption that both densities are independent necessarily leads to the conclusion that the volatility of the return process has a diffusion-like behavior, {\it i.e.} grows linearly with time, at any time scale. The hypothesis of independence is perhaps the simplest assumption one can make. However, it does not seem to be realistic since return variations  and their duration are certainly correlated, at least in many markets. We have therefore proposed a dependent model in which large return increments are infrequent. With this assumption the model predicts that the volatility should behave in an anomalous diffusive way at short times, something that is seen in some markets. 

The CTRW formalism allows us to obtain a closed expression for the joint Fourier-Laplace of the entire process which constitutes a convenient starting point for numerical analysis when no further analytical manipulations can be made. We have also obtain an asymptotic long-time expression for the characteristic function of the return process, this expression is valid as long as the pausing-time density possesses a finite first moment $\langle T\rangle$. 

We have applied the formalism in a study of the US dollar/Deutsche Mark future exchange market. We have used tic-by-tic data from 1993 to 1997. Data show that $\psi(t)$ and $h(x)$ are very well described by power-law densities ({\it cf} Eqs. (\ref{psifit}) and (\ref{hfit})). We have showed that (i) the tails of the return distribution follow a power-law with the same exponent as that of $h(x)$. (ii) The volatility has a diffusion-like behavior at long times and an anomalous diffusion-like behavior at short times. Both conclusion agree with experimental data. 

Let us finally mention that we have not been able to apply the CTRW formalism to market indices, such as the S\&P 500, since any index is an average of many prices and thus indices are recorded at fixed times. This contradicts the underlying assumption of the CTRW, {\it i.e., } that the time between successive changes is random. Therefore, the formalism presented herein is valid and applicable to single companies, currency exchange and commodities, while for market indices other formalisms, like the one presented in 
\cite{montero1}, are necessary.

\begin{acknowledgments}
This work has been supported in part by Direcci\'on General de
Investigaci\'on under contract No. BFM2000-0795 and by Generalitat de Catalunya under contract No. 2000 SGR-00023. We thank Mari\'an Bogu\~{n}\'a for many comments and suggestions to improve the manuscript.
\end{acknowledgments}


\begin{thebibliography}{999}
\bibitem{montrollweiss} E. W. Montroll and G. H. Weiss, J. Math. Phys. {\bf 6}, 167 (1965).
\bibitem{montroll2} E. W. Montroll and M. F. Shlesinger in {\it Nonequilibrium Phenomena II: From Stochastics to Hydrodynamics}, J. L. Lebowitz and E. W. Montroll eds. (North-Holland, Amsterdam, 1984).
\bibitem{weissllibre} G. H. Weiss, {\it Aspects and Applications of the Random Walk} (North-Holland, Amsterdam, 1994).
\bibitem{sher} H. Scher and E. W. Montroll, Phys Rev. B {\bf 12}, 2455 (1975). 
\bibitem{wpm} G. H. Weiss, J. M. Porr\`a, and J. Masoliver, Phys. Rev. E {\bf 55}, 6431 (1998); Opt. Commun. {\bf 146}, 268 (1998).
\bibitem{gandj} A. H. Gandjbakhche and G. H. Weiss, Phys. Rev. E {\bf 61}, 6958 (2000). 
\bibitem{berkowitz} B. Berkowitz and H. Scher, Phys. Rev. Lett. {\bf 79}, 4038 (1997). 
\bibitem{boguna} M. Boguna and A. Corral, Phys. Rev. Lett. {\bf 78}, 4950 (1997).
\bibitem{berezhkovskii} A. M. Berezhkovskii and G. Sutmann, Phys. Rev. E {\bf 65}, 060201 (2002).
\bibitem{gudowska} E. Gudowska-Nowak and K. Weron, Phys. Rev. E {\bf 65}, 011103 (2002).
\bibitem{balescu} R. Balescu, Phys. Rev. E {\bf 55}, 2465 (1997).
\bibitem{helmstetter} A. Helmstetter and D. Sornette, eprint cond-mat/0203505 (2002).
\bibitem{cootner} P.H. Cootner ed., {\it The Random Character of
Stock Market Prices} (MIT Press, Cambridge MA, 1964).
\bibitem{osborne} M. F. M. Osborne, Operations Research {\bf 7}, 145-173 (1959).
\bibitem{kendall} M.G. Kendall, J. Royal Stat. Soc. {\bf 96}, 11-25
(1953).
\bibitem{mandelbrot} B. Mandelbrot, J. Business {\bf 35}, 394-419 (1963).
\bibitem{mantegnastanley} R. N. Mantegna, and E.H. Stanley, Nature {\bf 376},
46-49 (1995).
\bibitem{montero1} J. Masoliver, M. Montero, and J.M. Porr\`{a}, Physica A
{\bf 283}, 559 (2000).
\bibitem{lukacs} This condition guarantees that $\tilde{\rho}(\omega,t)$ defined by 
Eq. (\ref{fform}) is a characteristic function. See E. Lukacs, {\it Characteristic Functions} (Griffin, London, 1970).
\bibitem{mos} W. Magnus, F. Oberhettinger, and R. P. Soni, {\it Formulas and Theorems for the Special Functions of Mathematical Physics} (Springer-Verlag, New York, 196
\bibitem{futures} Tic-by-tic data has been provided by the Futures Industries Institute 
(Washington, DC). The dollar/mark future exchange has been the object of numerous works 
(see, for instance, \cite{evertsz}).
\bibitem{evertsz} C. J. G. Evertsz, Fractals {\bf 3}, 609 (1995). 
\bibitem{plerou} V. Plerou, P. Gopikrishnan, L. A. Nunes Amaral, M. Meyer, and H. E. Stanley, Phys. Rev. E {\bf 60}, 6519 (1999).
\bibitem{gropi2} P. Gopikrishnan, V. Plerou, L. A. Nunes Amaral, M. Meyer, and H. E. Stanley, Phys. Rev. E {\bf 60}, 5305 (1999).
\bibitem{mainardi} F. Mainardi, M. Roberto, R. Gorenflo, and E. Scalas, Physica A 
{\bf 287}, 468 (2000). 
\bibitem{lighthill} M. J.Lighthill, {\it An Introduction to Fourier Analysis and Generalized Functions} (Cambridge University Press, Cambridge, 1980). 

\end{thebibliography}
\end{document}